\documentclass[aps,prd,,amsfonts,amssymb,amsmath,letterpaper,showkeys,10pt,reprint,showpacs]{revtex4-2}

\pdfoutput=1 

\usepackage{physics,graphicx}
\usepackage[colorlinks=true]{hyperref}

\newcommand{\gkg}[1]{\ensuremath{\langle #1 \rangle_{\text{gKG}}}}

\begin{document}


\title{Classical and quantum reconciliation of electromagnetic radiation: vector Unruh modes and zero-Rindler-energy photons}


\author{Felipe Portales-Oliva}
\email{felipe.portales@ufabc.edu.br}
\affiliation{Centro de Ci\^encias Naturais e Humanas, Universidade Federal do ABC, Avenida dos Estados, 5001, 09210-580, Santo Andr\'e, S\~ao Paulo, Brazil}

\author{Andr\'e G.\ S.\ Landulfo}
\email{andre.landulfo@ufabc.edu.br} 
\affiliation{Centro de Ci\^encias Naturais e Humanas, Universidade Federal do ABC, Avenida dos Estados, 5001, 09210-580, Santo Andr\'e, S\~ao Paulo, Brazil}


\begin{abstract}
A great deal of evidence has been mounting over the years showing a deep connection between acceleration, radiation, and the Unruh effect. Indeed, the fact that the Unruh effect can be codified in the Larmor radiation emitted by the charge was used to propose an experiment to experimentally confirm the existence of the Unruh thermal bath. However, such connection has two puzzling issues: {\bf (1)} how the quantum Unruh effect can be codified in the classical Larmor radiation and {\bf (2)} the fundamental role played by zero-Rindler-energy modes of the Unruh thermal bath in such a context. Here we generalize a recent analysis made for the scalar case to the more realistic case of Maxwell electrodynamics and settle these two puzzling issues. 
    \end{abstract}


\maketitle

\section{Introduction} \label{sec:introduction} 

Acceleration and radiation are deeply interconnected phenomena and their relationship has intrigued physicists for decades, dating back to the publication of Larmor's seminal
work~\cite{Larmor}. Although Larmor's formula played an instrumental role in establishing the relationship between acceleration and radiation, there were disagreements about its interpretation and validity (in particular in light of Einstein's equivalence principle, see, e.g., Refs.~\cite{Pauli.1918,FeynmanGravLects}). The works of Rohrlich~\cite{Rohrlich.1961,Rohrlich.1963} and Boulware~\cite{BOULWARE1980169} shed some light on the
classical aspects of this question. They found that radiation is not a covariant concept, as the perception of radiative phenomena is intrinsically associated to the observer's state of motion: if an inertial observer detects radiation coming from a uniformly accelerated charge, another observer in the coaccelerated
frame of the charge will not. However interesting and surprising that such classical results may be, it is in the context of quantum field theory (QFT) that the interplay between acceleration and radiation presents its most interesting hues.

In 1976, Unruh~\cite{PhysRevD.14.870} discovered that uniformly accelerated observers in the inertial vacuum perceive themselves immersed in a thermal bath of (Rindler) particles at a temperature
\begin{equation}
    T_{\mathrm{U}} = \frac{\hbar a}{2 \pi c k_{\mathrm{B}}},
    \label{eq:Unruh-temp}
\end{equation}
known as the Unruh temperature. The Unruh effect vindicated Fulling's previous discovery that the particle concept in quantum field theory is observer dependent~\cite{PhysRevD.7.2850} and provided the correct explanation to the observations by Davies~\cite{Davies_1975} that an accelerated observer registers radiation with a temperature proportional to his acceleration (see Ref.~\cite{RevModPhys.80.787} for a review on the Unruh effect). 

The connection between the Unruh effect and brems\-strah\-lung has been analyzed in Ref.~\cite{PhysRevD.46.3450} (see also  Refs.~\cite{PhysRevD.29.1047,PhysRevD.101.065012}), finding that the emission and absorption rate of zero-energy Rindler photons in the accelerated frame agrees with the emission rate of the
accelerated charge seen by the inertial perspective. Such an agreement can only be achieved by taking into account the existence of the thermal bath at temperature $ T_{\text{U}} $ in the accelerated frame. The deep link between the Unruh effect and brems\-strah\-lung has been strengthened by Ref.~\cite{PhysRevLett.118.161102}, where the authors found, rather surprisingly, that classical Larmor radiation codifies the purely quantum Unruh effect and provide an experimental procedure whose result can be directly interpreted in terms of the Unruh effect. More recently, in Ref.~\cite{PhysRevD.100.045020},  the authors analyzed the scalar version of the radiation emission problem to clarify {\bf (1)} the central role played by the zero-energy Rindler photons in the QFT calculations and {\bf (2)} how the classical Larmor radiation can codify the quantum Unruh effect. 
 
In this work, we expand Ref.~\cite{PhysRevD.100.045020} to the more realistic case of Maxwell electrodynamics. After extending the definition of Unruh modes to vector-valued solutions of the electromagnetic field equation (while retaining the important characteristics that made them useful for the scalar analysis) we
use them to expand the retarded potential associated with an uniformly accelerated charge and find the corresponding expansion amplitudes. With such an expansion, we define and compute the classical number of photons emitted by the charge and show the fundamental role played by the zero-energy Rindler photons of the Unruh thermal bath in such a process. Next, we proceed to the quantum analysis of the radiation emission by relating the field state in the asymptotic past and future, to find that not only the number of emitted photons will coincide with the classical case, but also the expectation values of the field and energy-momentum tensors agree with their classical counterparts. As in the classical case, it is made explicit how the radiation seen by the inertial observers can be traced-back to zero-energy Rindler photons absorbed/emitted from/to the Unruh thermal bath perceived in accelerated frame. 

The paper is structured as follows. In Sec.~\ref{sec:current-potential} we describe the details corresponding to the radiation emitted by an accelerated charge.
In Sec.~\ref{sec:Modes} we define vector Rindler and Unruh modes. Section~\ref{sec:decomposition-classical} is dedicated to finding the amplitudes associated with the Unruh mode decomposition of the classical field, while  Sec.~\ref{secdecomposition-quantum} is reserved to the quantum calculations with a classical
source. Some final remarks and discussion are presented in Sec.~\ref{sec:conclusions}. We work with a 4-dimensional spacetime with metric signature $(+,-,-,-) $, along with Heaviside-Lorentz units for the electromagnetic quantities, and set $ \hbar = c = 1 $ throughout the paper.

\section{The 4-current and 4-potential from a uniformly accelerated charge}
\label{sec:current-potential} 

The Lagrangian density for the electromagnetic field that allows for the field quantization in a globally hyperbolic spacetime $(\mathcal{M}, g_{ab})$ is given by 
\begin{equation}
    \mathcal{L} 
    = 
    -
    \sqrt{-g}
    \qty(
        \frac{1}{4} F^{ab} F_{ab}
        +
        \frac{1}{2\alpha} (\nabla_a A^a)^2
        +
        j_a A^a
    ),
    \label{eq:EM-lagrangian-density}
\end{equation}
where $F_{ab}\equiv 2\nabla_{[a}A_{b]}$, $A_a$ is the 4-potential, $ j_a $ is the 4-current source of the electromagnetic field, $ \nabla_a $ corresponds to the covariant derivative compatible with the Lorentzian metric $g_{ab}$, and $g$ indicates the determinant of $g_{ab}$ in some arbitrary coordinate system~\cite{Wald84}. The field equation for a Ricci-flat ($ {R}_{ab} = 0 $) spacetime in the Feynman gauge, $ \alpha = 1 $, is given by
\begin{equation}
    \nabla_b \nabla^b A_a 
    = 
    j_a,
    \label{eq:field-eq}
\end{equation}
which can also be  achieved by imposing
\begin{equation}
    \nabla_a A^a = 0,
    \label{eq:lorenz-condition}
\end{equation}
known as the Lorenz condition, \emph{a priori}.

\begin{figure}[h] 
	\centering 
	\includegraphics[width=8.6cm]{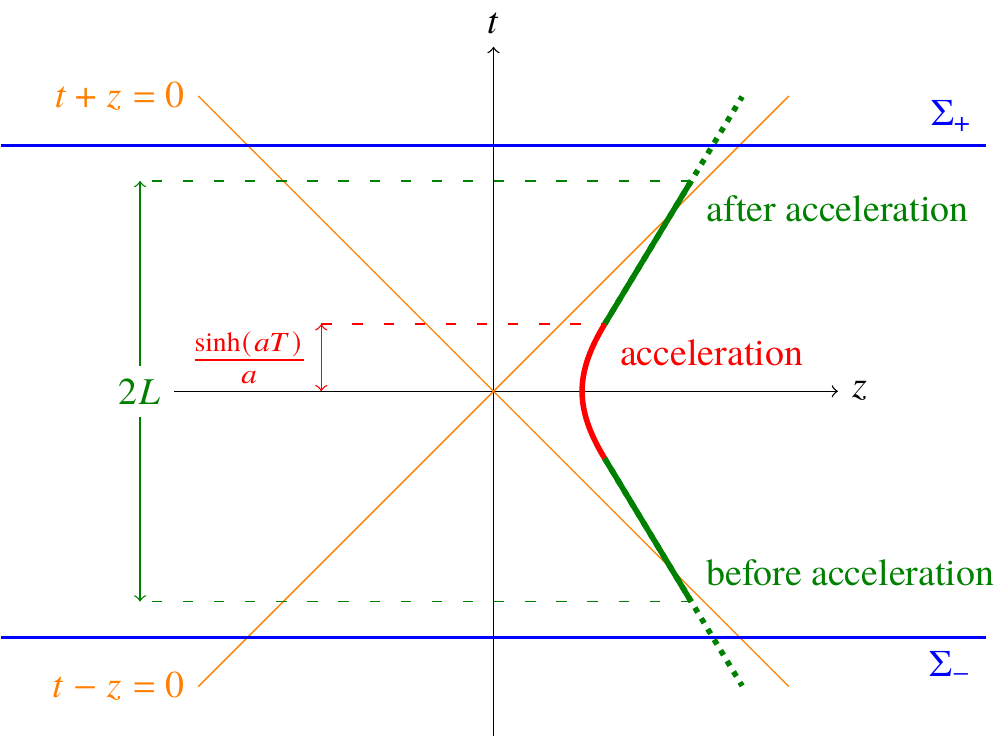}
  \caption{Spacetime diagram describing the worldline of an initially inertial charge $q$ which is uniformly accelerated for a finite proper time $2T$ and then becomes inertial again. The slices $\Sigma_-$ and $\Sigma_+$ are Cauchy surfaces to the past and future, respectively, of the support of the compactified 4-current $j^a_L$}
  \label{fig:trajectory} 
\end{figure}

Let us consider an initially inertial charge $ q $ in Min\-kow\-ski spacetime $(\mathbb{R}^4, \eta_{ab})$ which is accelerated with constant proper acceleration $a$ for a finite proper time  $2T$ and then becomes inertial again. Here, $\eta_{ab}$ indicates the Minkowski metric whose line element in usual inertial Minkowski
coordinates $(t,x,y,z)$ takes the form
\begin{equation}
   \dd s^2 = \dd t^2 -\dd x^2- \dd y^2 - \dd z^2 .
    \label{eq:Minkds2}
\end{equation}
In such coordinates, the worldline of the charge is given by (see Fig.~\ref{fig:trajectory})
\begin{widetext}
\begin{equation}
    z^a (t) =
    \begin{cases}
        \boldsymbol( t,0,0, a^{-1}\cosh(aT) - \tanh(aT) [t + a^{-1} \sinh(aT)] \boldsymbol) 
            & \text{if $ t \leq - a^{-1} \sinh(aT) $}, \\
        (t,0,0,\sqrt{a^{-2}+t^2}) 
            & \text{if $ |t| < a^{-1} \sinh(aT) $}, \\
        \boldsymbol( t,0,0, a^{-1}\cosh(aT) + \tanh(aT) [t - a^{-1} \sinh(aT)] \boldsymbol) 
            & \text{if $ t \geq a^{-1} \sinh(aT) $},
    \end{cases}
    \label{eq:trajectory-charge-conserved-constant-vel}
\end{equation}
and the 4-current associated with this worldline is given by
\begin{equation}
    j^a (x)
    =
    \begin{cases}
        j_{I}^a(x) 
            & \text{if $ |t| \geq a^{-1} \sinh(aT) $},
        \\
        j_{A}^a (x)     & \text{if $ |t| < a^{-1} \sinh(aT) $}.
    \end{cases}
    \label{eq:conserved-4-current}
\end{equation}
Here, $j^a_I$ is the current associated with the inertial part of the motion and has components
\begin{subequations} \label{eq:conserved-4-current-smooth}
    \begin{gather}
        j_I^t (x)
        =
        \begin{cases}
            q \cosh(aT)\, \delta^2(\mathbf{x}_\perp) \, \Delta_+(z,t) 
            & \text{if $ t \leq -a^{-1} \sinh(aT) $},
            \\
            q \cosh(aT) \, \delta^2(\mathbf{x}_\perp) \, \Delta_-(z,t)  
                & \text{if $ t \geq a^{-1} \sinh(aT) $},
        \end{cases}
        \label{eq:conserved-4-current-smooth-time}
        \\
        j_I^z (x)
        =
        \begin{cases}
            -q \sinh(aT) \, \delta^2(\mathbf{x}_\perp) \, \Delta_+(z,t)
                & \text{if $ t \leq -a^{-1} \sinh(aT) $},
            \\
            q \sinh(aT) \, \delta^2(\mathbf{x}_\perp) \, \Delta_-(z,t) 
                & \text{if $ t \geq a^{-1} \sinh(aT) $},
        \end{cases}
        \label{eq:conserved-4-current-smooth-space}
    \end{gather}
\end{subequations}
along with $j_I^x(x)=j_I^y(x)=0$. Here we have defined the auxiliary functions
\begin{equation}
    \Delta_\pm(z,t) 
   \equiv  
    \delta \boldsymbol(z-a^{-1}\sech(aT) \pm t\tanh(aT)  \boldsymbol).
    \label{eq:auxiliary-function}
\end{equation}
\end{widetext}
The current $j^a_A$ describes the uniformly accelerated part of the motion. In the right Rindler wedge (RRW), i.e. the spacetime region defined by $z>|t|$, $ j_A^a (x) $ can be cast in a simple form by using Rindler coordinates $(\lambda, \xi, x, y) $,
where $ x $ and $ y $ are left unaltered while the other two inertial coordinates are written as
\begin{align}\label{eq:right-Rindler-coords}
    t &= a^{-1} e^{a\xi} \sinh(a\lambda),
    &
    z &= a^{-1} e^{a\xi} \cosh(a\lambda),
\end{align}
yielding 
\begin{equation}
    j_A^a = 
        q \, \delta(\xi) \, \delta^2(\vb{x}_\perp) \, \theta(T-|\lambda|)
        (\partial_\lambda)^a.
        \label{jA}
\end{equation}
Here, $ \delta $ is the Dirac delta distribution, $ \theta $ is the Heaviside step function, and $ \vb{x}_\perp \equiv  (x , y ) \in \mathbb{R}^2 $. We have included the step function to reinforce the fact that this part of the current is limited to the region defined by $ |t| < a^{-1} \sinh(aT) $.
Notice the trajectory \eqref{eq:trajectory-charge-conserved-constant-vel} is not constrained to the RRW (like an infinitely accelerated particle would): before the acceleration the charge moves in the Contracting Degenerate Kasner Universe (CDKU, $t<-|z|$), it then crosses the Killing horizon $t+z=0$ to the RRW, and after the acceleration it crosses the surface $t-z=0$ into the Expanding Degenerate Kasner Universe (EDKU, $t>|z|$).

As we are interested in describing radiation, let us recall from
the classical theory that the energy radiated by a point charge over a unit solid angle per emission time is given by~\cite{zangwill2013modern}
\begin{align}
    \frac{\dd^2 H}{\dd \Omega \, \dd t_{\text{ret}} } 
    &= 
    \lim_{R\to\infty}
    \qty(1 - \vb{v}(t_{\text{ret}}) \cdot \vb{n}(t_{\text{ret}}) )
    R^2 \vb{S}(t) \cdot \vb{n}(t_{\text{ret}})
    \nonumber \\
    &=
    \frac{q^2}{16 \pi^2}
    \qty[ 
        \frac{
            |
                \vb{n} \times [ ( \vb{n}-\vb{v} ) \times \vb{a} ]
            |^2
        }{
            ( 1-\vb{v} \cdot \vb{n} )^5
        }
    ]_{\text{ret}},
    \label{eq:rad-power-by-solid-angle}
\end{align}
where $ \vb{n} $ is a unitary vector that points from the point of emission to the point of evaluation, $ \vb{v} $ is the velocity of the charge, and $ \vb a $ is its acceleration.  We can see from Eq.~(\ref{eq:rad-power-by-solid-angle}) that the only contribution to the radiated energy comes from the parts of the trajectory where acceleration is nonzero. Indeed, by using the worldline~\eqref{eq:trajectory-charge-conserved-constant-vel} of our charge together with  $\vb{n} = ( \sin\theta\cos\varphi,\sin\theta\sin\varphi ,\cos\theta ) $, where the angles $ \theta $ and $\varphi $ are defined from the point of emission, Eq.~\eqref{eq:rad-power-by-solid-angle} reduces to 
\begin{equation}
    \frac{\dd^2 H}{\dd \Omega \, \dd t_{\text{ret}} }
    =
    \left.
    \frac{
        (16 \pi^2)^{-1} a^2 q^2 \sin^2\theta 
    }{
        \sqrt{ 1 + a^2 t^2} 
        ( \sqrt{ 1 + a^2 t^2 } - a t \cos\theta)^5
    }
    \right|_{\text{ret}}
    \label{eq:rad-power-by-solid-angle-aux1}
\end{equation}
if $ |t_\text{ret}| < a^{-1} \sinh(aT)  $, and to zero otherwise.

The Lié\-nard-Wie\-chert potential produced by the current~(\ref{eq:conserved-4-current}) is given by~\cite{zangwill2013modern,jackson}
\begin{equation}
    R j^a (x) 
    =
    \frac{q}{2\pi}
    \int_{-\infty}^{\infty}
    \dd\tau
    \,
    {\dv{z^a}{\tau}}  
    \theta \boldsymbol( x^0 - z^0(\tau) \boldsymbol)
    \,
    \delta\boldsymbol( [x-z(\tau)]^2 \boldsymbol)
    ,
    \label{eq:lienard-wiechert}
\end{equation}
where $\tau$ is the charge's proper-time. Born~\cite{born1909theory} used this expression for the case where the charge is uniformly accelerated for an infinite amount of time (result which was further
discussed and refined in works like~\cite{fulton1960classical,BOULWARE1980169}) yielding 
\begin{subequations} \label{eq:Born-potential-inertial}
    \begin{gather}
        R j^t 
        =
        \frac{q}{4\pi(t^2-z^2)}
        \qty[ 
            \frac{
                a z  (a^{-2} - t^2 + r^2 )
            }{
                    2 \rho_0(x)
            }   
            - t
        ]
        \theta(t+z)
        ,
        \\
        Rj^z 
        =
        \frac{q}{4\pi(t^2-z^2)}
        \qty[ 
            \frac{
                a t  (a^{-2} - t^2 + r^2 )
            }{
                    2 \rho_0(x)
            }
            - z
        ]
        \theta(t+z)
        ,
        \\
        R j^x = Rj^y = 0,
    \end{gather}
\end{subequations}
where $ r^2 = x^2 + y^2 + z^2 $ and 
\begin{equation}
    \rho_0(x) \equiv \frac{a}{2} \sqrt{
        \frac{4}{a^2}(t^2 - z^2)
        + 
        \left[ \frac{1}{a^2} - t^2 + r^2 \right]^2 
    }.
    \label{rho0}
\end{equation} 
This solution codifies all the information about the charge $q$ following the worldline~(\ref{eq:trajectory-charge-conserved-constant-vel}) in the limit
$T\to\infty$. Later in the paper, we will show how this solution is built entirely from zero-energy Rindler modes. 

For the sake of some calculations, it will be useful to consider the following compactified version of the current~(\ref{eq:conserved-4-current}):
\begin{equation}\label{eq:jL}
    j_L^a\equiv  \theta(L-|t|)  j^a,
\end{equation}
which will enable us to use some useful mathematical identities. Here, we have introduced a \emph{compactification parameter} $ L > a^{-1} \sinh(aT) $. We note that $j^a_L$ is not locally conserved  as 
\begin{align}
    \nabla_a j_L^a
    &= 
    j^t \, \delta(L-|t|) 
    \neq 0,
\end{align}
where we used the current $j^a$ defined in Eq.~\eqref{eq:conserved-4-current}, which does satisfy $ \nabla_a j^a =0 $.
However as $ L $ is a free parameter, we can compute the quantities of interest using this current of compact support and then recover the physically relevant results, where charge is locally conserved, by taking the limit $ L \to \infty $.  As our aim is to study the radiation emitted by a charge accelerated during a finite amount of its proper-time, and the inertial portions of the motion of this charge will not contribute to the radiation emitted (nor detected), the analysis of the radiation using $j^a_L$ (eventually taking $L\rightarrow \infty$) will be enough for our purposes. It is important to note that we will also present the corresponding
correction terms coming from the inertial parts of the motion and the compactification, to guarantee that it does not contribute when the limit $L\rightarrow\infty$ is taken.

\section{Rindler and Unruh modes}\label{sec:Modes}

The main idea behind this study is to decompose both the classical and quantum electromagnetic 4-potential in terms of vector Unruh modes; the definition of which is based on Rindler modes that appear on the description of a scalar field on both Rindler wedges. In the following we recall the discussion for the scalar case as it is fundamental to our procedure. 

Let us consider a free scalar field $\phi$ in $(\mathbb{R}^4, \eta_{ab})$ satisfying the Klein-Gordon (KG) equation
\begin{equation}
    \nabla_a \nabla^a \phi = 0.
    \label{eq:KGeq}
\end{equation}
As the RRW is a globally-hyperbolic static spacetime on its own right, with time-like Killing field $(\partial_\lambda)^a$, one can define the set of positive-frequency right Rindler modes $\{ v^{R}_{\omega\vb{k}_\perp} \}$ which are solutions of Eq.~\eqref{eq:KGeq} vanishing in the left Rindler Wedge (LRW), i.e., the spacetime region defined by $t<-|z|$, and taking the form
\begin{equation}
    v^{R}_{\omega\vb{k}_\perp} 
    =
    \sqrt{
        \frac{
            \sinh(\pi\omega/a)
        }{
            4 \pi^4 a
        }
    }
    K_{i\omega/a} \qty(\frac{k_\perp e^{a\xi}}{a})
    e^{i  \vb{k}_\perp\cdot\vb{x}_\perp} e^{-i\omega\lambda },
    \label{eq:right-rindler-modes}
\end{equation}
in the RRW using Rindler coordinates. Here, $ K_\nu(z) $ is the modified Bessel function of the second kind, $\vb{k}_\perp \equiv (k_x, k_y)  \in \mathbb{R}^2
- \{0\}$, and $ \omega
\in \mathbb{R}^+$. We can also define left Rindler modes $v^{L}_{\omega\vb{k}_\perp} $ by means of the relation 
\begin{equation}
    v^L_{\omega {\bf k}_\perp}(t,{\bf x}_\perp,z)\equiv {v^{R*} _{\omega {\bf k}_\perp}}(-t,{\bf x}_\perp,-z).
     \label{eq:left-rindler-modes}
\end{equation}
Hence, they vanish in the RRW and take the form~(\ref{eq:right-rindler-modes}) in Rindler coordinates covering the LRW. The set of modes $\{v^{R}_{\omega\vb{k}_\perp}, v^{L}_{\omega\vb{k}_\perp}\}$, together with their Hermitian conjugates, forms a complete set of solutions of the Klein-Gordon equation in Minkowski spacetime~\cite{RevModPhys.80.787}. 

By using the Rindler modes~(\ref{eq:right-rindler-modes}) and ~(\ref{eq:left-rindler-modes}), we can define another suitable set of solutions of the KG equation, the so-called {\em Unruh Modes} $\{w_{\omega\vb{k}_\perp}^{1},  w_{\omega\vb{k}_\perp}^{2} \}$. They are defined as  
\begin{subequations} \label{eq:unruh-modes}
    \begin{align}
        w_{\omega\vb{k}_\perp}^{1} 
        &\equiv
            \frac{v^{R}_{\omega\vb{k}_\perp} 
                + e^{-\pi\omega/a} v^{L*}_{\omega\, -\vb{k}_\perp} 
                }
                { \sqrt{ 1-e^{-2\pi\omega/a} } },
        \\
        w_{\omega\vb{k}_\perp}^{2} 
        &\equiv 
            \frac{v^{L}_{\omega\vb{k}_\perp} 
                + e^{-\pi\omega/a} v^{R*}_{\omega\, -\vb{k}_\perp} 
                }
                { \sqrt{ 1-e^{-2\pi\omega/a} } }.
    \end{align}
\end{subequations}
and, although they are labeled by the Rindler energy $\omega$ and transverse momentum ${\bf k}_\perp$, they are positive-frequency with respect to the {\em inertial time} $t$ and form (together with their Hermitian conjugate) a complete set of orthonormal solutions of the KG equation. This makes them  suitable to investigate the relation between radiation seen by inertial observers and the physics of uniformly accelerated observers.

Having defined the scalar Unruh modes, we can now turn our attention to the electromagnetic case. The solutions of the homogeneous electromagnetic field equation in $(\mathbb{R}^4, \eta_{ab})$,
\begin{equation}
    \nabla_b\nabla^b A_a = 0,
    \label{eq:homogeneous-EM}
\end{equation}
can be decomposed in terms of 4 independent  polarization modes on the RRW  given by~\cite{PhysRevD.46.3450}
\begin{subequations} \label{eq:right-rindler-polarization-modes}
    \begin{gather}
        V_{{\omega\vb{k}_\perp}\ a}^{R(1)} 
        = 
        \frac{1}{k_\perp} (0,0,k_y v^{R}_{\omega\vb{k}_\perp},-k_xv^{R}_{\omega\vb{k}_\perp}),
        \\
        V_{{\omega\vb{k}_\perp}\ a}^{R(2)} 
        = 
        \frac{1}{k_\perp} (\partial_\xi v^{R}_{\omega\vb{k}_\perp}, \partial_\lambda v^{R}_{\omega\vb{k}_\perp},0,0),
        \\
        V_{{\omega\vb{k}_\perp}\ a}^{R(G)} 
        = 
        \frac{1}{k_\perp} \nabla_a v^{R}_{\omega\vb{k}_\perp},
        \\
        V_{{\omega\vb{k}_\perp}\ a}^{R(L)} 
        = 
        \frac{1}{k_\perp} (0,0,k_x v^{R}_{\omega\vb{k}_\perp},k_y v^{R}_{\omega\vb{k}_\perp}),
    \end{gather}
\end{subequations}
where $v^{R}_{\omega\vb{k}_\perp}$ are the scalar right Rindler modes given in Eq.~(\ref{eq:right-rindler-modes}). These have been selected in such a way that they are orthonormalized with respect to the generalized Klein-Gordon inner product 
\begin{equation}
    \gkg{A^{(1)},A^{(2)}}
    =
    \int_{\Sigma}
    \dd\Sigma_a 
    \,
    \Xi^a [A^{(1)},A^{(2)}]
    \label{eq:gen-KG-prod}
\end{equation}
between two modes, $ A^{(1)},A^{(2)},$ of the electromagnetic field. Here, the integration is done over any Cauchy surface $ \Sigma $, with proper-vector-valued-volume element $\dd \Sigma_a$, and  conserved current $ \Xi^a [A^{(1)},A^{(2)}] $  given by 
\begin{equation}
    \Xi^a [A^{(1)},A^{(2)}]
    \equiv
    \frac{i}{\sqrt{-g}} \qty(
        A_b^{(1)*} \pi^{(2)ab}
        -
        A_b^{(2)} \pi^{(1)ab*}
    ),
    \label{eq:genKG-EM-current}
\end{equation}
where we have used the generalized momenta of the electromagnetic potential defined as  $ \pi^{ab} \equiv {\partial\mathcal{L}}/{\partial(\partial_a
A_b)} $. On the Feynman gauge these are computed explicitly yielding 
\begin{equation}
    \pi^{ab}
    =
    \sqrt{-g}
    \qty(
        \nabla^b A^a - \nabla^a A^b - g^{ab} \nabla_c A^c
    ).
    \label{eq:gen-momentum-4potential}
\end{equation}
The left electromagnetic Rindler modes are defined, as in the scalar case, as 
\begin{equation}
 V_{{\omega\vb{k}_\perp}\ a}^{L(\kappa)}(t,{\bf x}_\perp,z)\equiv V_{{\omega\vb{k}_\perp}\ a}^{R(\kappa)* }(-t,{\bf x}_\perp,-z), 
\end{equation}
with $\kappa=1,2, G, L$. 
Here, the modes (both left and right) labeled with $\kappa = 1,2$ are physical modes, while the ones labeled with $\kappa = G,L$ are non-physical, as they are pure gauge and do not satisfy the Lorenz condition, respectively.

By making use of  $V_{{\omega\vb{k}_\perp}\ a}^{R(\kappa)}$ and $ V_{{\omega\vb{k}_\perp}\ a}^{L(\kappa)}$, we can extend the definition of the Unruh modes to the electromagnetic case, in analogous fashion to the definitions of Eq.~\eqref{eq:unruh-modes}
\begin{subequations} \label{eq:EM-unruh-modes}
    \begin{align} 
        W_{\omega\vb{k}_\perp \, b}^{1 (\kappa)} 
        &\equiv
        \frac{
            V^{R(\kappa)}_{\omega\vb{k}_\perp \, b} 
            + e^{-\pi\omega/a} V^{L(\kappa)*}_{\omega\, -\vb{k}_\perp\, b} 
        }{ 
            \sqrt{ 1-e^{-2\pi\omega/a} } },
        \\ 
        W_{\omega\vb{k}_\perp \, b}^{2 (\kappa)} 
        &\equiv 
        \frac{
            V^{L(\kappa)}_{\omega\vb{k}_\perp \, b} + e^{-\pi\omega/a} V^{R(\kappa)*}_{\omega\, -\vb{k}_\perp \, b} 
        }{ 
            \sqrt{ 1-e^{-2\pi\omega/a} } 
        }.
    \end{align}    
\end{subequations}
These, by using Eqs.~(\ref{eq:unruh-modes}) and~(\ref{eq:right-rindler-polarization-modes}), can be expressed directly in terms of the scalar Unruh modes as
\begin{subequations} \label{eq:explicit-EM-unruh-modes}
    \begin{gather}
        W_{\omega\vb{k}_\perp\ a}^{\sigma(1)} 
        = 
            \frac{1}{k_\perp} 
            (
                0 ,
                k_y w^{\sigma}_{\omega\vb{k}_\perp} ,
                -k_x w^{\sigma}_{\omega\vb{k}_\perp} ,
                0
            ),\label{eq:explicit-EM-unruh-modes-a}
        \\
        W_{\omega\vb{k}_\perp\ a}^{\sigma(2)} 
        =
            \frac{1}{k_\perp} 
            (
                \partial_z w^{\sigma}_{\omega\vb{k}_\perp}, 
                0,
                0,
                \partial_t w^{\sigma}_{\omega\vb{k}_\perp}
            ),
        \label{eq:explicit-EM-unruh-modes-b}
        \\
        W_{\omega\vb{k}_\perp\ a}^{\sigma(G)} 
        = 
            \frac{1}{k_\perp} \nabla_a w^{\sigma}_{\omega\vb{k}_\perp},
        \\
        W_{\omega\vb{k}_\perp\ a}^{\sigma(L)} 
        = 
            \frac{1}{k_\perp} 
            (
                0,
                k_x w^{\sigma}_{\omega\vb{k}_\perp},
                k_y w^{\sigma}_{\omega\vb{k}_\perp},
                0    
            ).\label{eq:explicit-EM-unruh-modes-d}
    \end{gather}
\end{subequations}
Hence, they satisfy 
\begin{multline}
    \gkg{
        W_{\omega\vb{k}_\perp}^{\sigma(\kappa)}
        ,
        W_{\omega'\vb{k}'_\perp}^{\sigma'(\kappa')}
    }\\
    =
    \delta_{\sigma\sigma'}
    \delta_{\kappa\kappa'}
    \delta(\omega-\omega')
    \delta^2(\vb{k}_\perp-\vb{k}'_\perp),
    \label{eq:orthogonality-EM-Unruh-modes}
\end{multline}
and form (together with they Hermitian conjugate) a complete set of solutions for the homogeneous electromagnetic field equation \eqref{eq:homogeneous-EM} which are positive-frequency with respect to the inertial time $t$.


\section{Unruh mode decomposition of the classical retarded potential}\label{sec:decomposition-classical}

\begin{figure}[hbt] 
	\centering 
	\includegraphics[width=8.6cm]{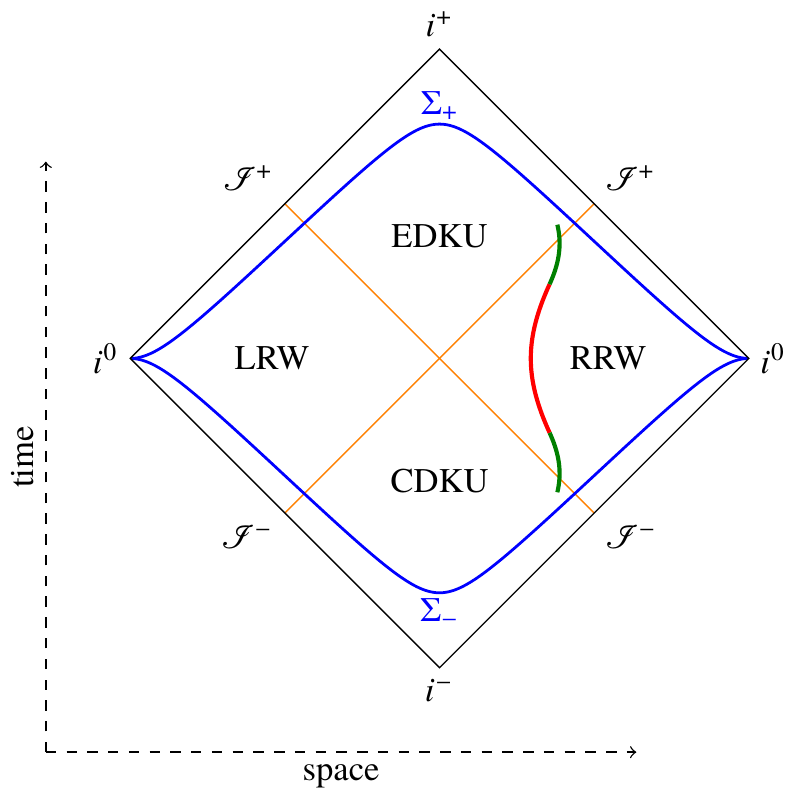}
  \caption{Conformal diagram of our set-up. The red portion of the worldline displays the support of the accelerated part of the compactified current $j^a_L$, while the support of the inertial parts are shown in green. The orange lines depicts the Killing horizons and the blue ones the Cauchy surfaces.}
  \label{fig:diagram} 
\end{figure}

Let us consider two Cauchy surfaces,  $ \Sigma_+ $ and $ \Sigma_- $, in Minkowski spacetime with $\Sigma_+ \subset I^+\left(\Sigma_-\right)$ and $\Sigma_\pm
\subset \mathbb{R}^4-{\rm supp} (j_L^a)$ (see Fig.~\ref{fig:diagram}). Here,  $I^+\left(A\right)$ denotes  the chronological future of a subset $A\subset
\mathbb{R}^4$.

The advanced and retarded solution of the field equation~\eqref{eq:field-eq} with source $j^a_L$  are given by 
\begin{subequations} \label{eq:adv-ret-solutions}
    \begin{gather}
        Aj_L^a(x) 
        = 
        \int_{\mathbb{R}^4} \dd^4x' \sqrt{-g'} 
            G_{\text{adv}}(x,x') j_L^a(x'), 
        \label{eq:adv-solution}\\
        Rj_L^a(x) 
        = 
        \int_{\mathbb{R}^4} \dd^4x' \sqrt{-g'} 
            G_{\text{ret}}(x,x') j_L^a(x'),
        \label{eq:ret-solution}
    \end{gather}
\end{subequations}
respectively, where $ G_{\text{adv}}$ and $ G_{\text{ret}} $ are the advanced and retarded Green's functions for the electromagnetic field~\cite{jackson}. Given our choice for Cauchy surfaces, we find that 
\begin{equation} 
        Rj_L^a(x) = -Ej_L^a(x) \quad \forall x\in\Sigma_+,
        \label{eq:ret-field-future}
\end{equation}
where we have defined $ Ej_L^a \equiv  Aj_L^a - Rj_L^a $ and we have used that $Aj^a_L$ vanishes in $\mathbb{R}^4-J^-\left({\rm supp} j^a_L\right)$.

In order to analyze the role played by the Unruh thermal bath (with particular interest in the zero-energy Rindler modes) in building the radiation emitted by the charge as seen by inertial observers, it will be important to decompose $Rj^a_L$ in terms of Unruh modes~(\ref{eq:explicit-EM-unruh-modes}). To this end, we will use our compactified current $j^a_L$ to compute $Rj^a_L$ and take the limit $L\rightarrow \infty$ to recover the physical current case. Such a task will be greatly simplified by using the identity [proven in Appendix~\ref{appA}]
\begin{equation}
    \gkg{\tilde{A},E\tilde{j}} 
    = 
    -i \int_{\mathcal{M}} \dd^4 x \sqrt{-g} \, \tilde{A}_a^* \tilde{j}^a,
    \label{eq:identity}
\end{equation}
which is valid for any Ricci-flat ($R_{ab}=0$) globally-hyperbolic spacetime $(\mathcal{M}, g_{ab} )$, with $\tilde{A}^a$ being any solution of the {\em
homogeneous} electromagnetic field equation~(\ref{eq:homogeneous-EM}) and $\tilde{j}^a$ any compact-support current. 

Let us now  expand Eq.~(\ref{eq:ret-field-future}) in terms of Unruh modes on $ \Sigma_+ $ as
\begin{multline}
    {Rj_L}_a 
    =
    -
    \sum_{\sigma,\kappa}
    \int_0^\infty \dd\omega
    \int_{\mathbb{R}^2} \dd^2\vb{k}_\perp
    \\
    \qty(
        \gkg{W_{\omega\vb{k}_\perp}^{\sigma(\kappa)},Ej_L} 
            \, 
            W_{\omega\vb{k}_\perp\ a}^{\sigma(\kappa)}
        +
        \text{c.c.}
    ),
    \label{eq:Unruh-mode-expansion}
\end{multline}
where $ \text{c.c.} $ represents the complex conjugate of the previous expression. The coefficients $\gkg{W_{\omega\vb{k}_\perp}^{\sigma(\kappa)} , Ej_L } $ can be computed  using Eq.~\eqref{eq:identity}. We immediately find that $\gkg{W_{\omega\vb{k}_\perp}^{\sigma(1)},Ej_L} = \gkg{W_{\omega\vb{k}_\perp}^{\sigma(L)},Ej_L} = 0 $, as the current does not couple with the corresponding modes.  As the support of $j^a_L$ is contained in the RRW (even on the limit $ T\to\infty $), by  using Eqs.~(\ref{eq:conserved-4-current}) and~(\ref{eq:jL}) we can cast the other coefficients as 
\begin{subequations}\label{eq:expansion-coefficients-finite-time}
    \begin{multline}
        \gkg{W_{\omega\vb{k}_\perp}^{1(2)},Ej_L}
        =
        - \frac{ i q e^{\pi\omega/2a} \sin(\omega T)}{\sqrt{ 2 \pi^4 a }\omega}
        K'_{i\omega/a} (k_\perp/a) \\
        + \mathcal{I}^{1(2)}_{\omega {\bf k}_\perp}(a, T, L)
        ,
    \end{multline}
    \begin{multline}
        \gkg{W_{\omega\vb{k}_\perp}^{1(G)},Ej_L} 
        =
        \frac{ q \sin(\omega T) e^{ \pi\omega/2a }}
            {\sqrt{ 2 \pi^4 a}{k_\perp}}
        K_{i\omega/a} (k_\perp/a) \\
        + \mathcal{I}^{1(G)}_{\omega {\bf k}_\perp}(a, T, L)
        ,
    \end{multline}
    \begin{multline}
        \gkg{W_{\omega\vb{k}_\perp}^{2(2)},Ej_L}
        =
        -\frac{ i q \sin(\omega T) }
            {\sqrt{ 2\pi^4 a e^{\pi\omega/a} }\omega}
        K'_{i\omega/a} (k_\perp/a) \\
        + \mathcal{I}^{2(2)}_{\omega {\bf k}_\perp}(a, T, L)
        ,
    \end{multline}
    and
    \begin{multline}
    \gkg{W_{\omega\vb{k}_\perp}^{2(G)},Ej_L}
        =
        -\frac{q \sin(\omega T)}
            {\sqrt{ 2 \pi^4 a e^{\pi\omega/a} }{k_\perp}}
        K_{i\omega/a} (k_\perp/a) \\
        + \mathcal{I}^{2(G)}_{\omega {\bf k}_\perp}(a, T, L),
    \end{multline}
\end{subequations}
where we have used the prime to denote differentiation with respect to the argument. Here, we have written the amplitudes in Eq.~(\ref{eq:expansion-coefficients-finite-time}) separating two different contributions: the first one coming from the uniformly accelerated part of the current $j^a_L$ and the second one, 
\begin{gather}
    \mathcal{I}^{\sigma (\kappa)}_{\omega\mathbf{k}_\perp} 
    \equiv  
    - i q \left[
        \cosh(aT) \mathcal{A}^{\sigma(\kappa)}_{\omega\mathbf{k}_\perp} 
        + 
        \sinh(aT) \mathcal{B}^{\sigma(\kappa)}_{\omega\mathbf{k}_\perp}
    \right],
    \label{eq:correction-smooth-mode2}
\end{gather}
with 
\begin{eqnarray}
    \mathcal{A}^{\sigma(\kappa)}_{\omega\mathbf{k}_\perp}
    &\equiv & 
    \int_{a^{-1} \sinh(a T)}^{L} \mathrm{d} t \int_{\mathbb{R}}\mathrm{d} z\left[
        W_{\omega\mathbf{k}_\perp}^{\sigma(\kappa) * } {}_{t}(t,0,0,z) \Delta_-(z,t)\right. \nonumber  \\ 
    &+&  \left. W_{\omega\mathbf{k}_\perp}^{\sigma(\kappa) *}{}_{t}(-t,0,0,z) \Delta_+(z,-t) \right],
    \label{eq:A.correction}
    \\
    \mathcal{B}^{\sigma(\kappa)}_{\omega\mathbf{k}_\perp}
    &\equiv & \int_{a^{-1} \sinh(a T)}^{L} \mathrm{d} t \int_{\mathbb{R}}\mathrm{d} z \left[
        W_{\omega\mathbf{k}_\perp}^{\sigma(\kappa) * } {}_{z}
            (t,0,0,z) \Delta_-(z,t)\right.\nonumber  \\ 
            &-&  \left. W_{\omega\mathbf{k}_\perp}^{\sigma(\kappa) *}{}_{z} (-t,0,0,z) \Delta_+(z,-t)\right], 
    \label{eq:B.correction}
\end{eqnarray}
coming from the inertial part of $j^a_L$. 
Now, after computing these 
integrals, we take the limit $L\to\infty $ to obtain the complete result for our physical current, i.e., the current given in Eq.~\eqref{eq:conserved-4-current} which satisfy  $ \nabla_a j^a = 0 $. In Appendix~\ref{appB2} we present the details of such calculation. 
  
Having determined the amplitudes for our physical current, let us now find their form in the limit where the charge accelerates for an infinite proper-time, i.e., $T\rightarrow \infty$. By using 
\begin{equation}
    \lim_{T \to \infty} \frac{\sin(\omega T)}{\omega} = \pi \delta(\omega),
    \label{eq:limit}
\end{equation}
the identity $ K'_\nu(z) = \nu K_\nu(z)/z - K_{\nu+1}(z) $ for the modified Bessel functions of the second kind, and the fact [proven in Appendix~\ref{appB2}] that $\mathcal{I}^{\sigma (2)}_{\omega\mathbf{k}_\perp}$  and $ \mathcal{I}^{\sigma (G)}_{\omega\mathbf{k}_\perp}$ vanish for $T\rightarrow \infty,$  we find that the amplitudes of Eq.~(\ref{eq:expansion-coefficients-finite-time}) can be cast as
\begin{subequations} 
    \begin{align}
        \gkg{W_{\omega\vb{k}_\perp}^{1(2)},Ej} 
        &= \gkg{W_{\omega\vb{k}_\perp}^{2(2)},Ej}
        \nonumber \\
        &= \frac{i q}{\sqrt{2a}\pi} K_1\!\left(\frac{k_\perp}{a}\!\right)\!\delta(\omega),  \label{eq:expansion-coefficients-infinite-timea} \\
       \gkg{W_{\omega\vb{k}_\perp}^{1(G)},Ej} &=\gkg{W_{\omega\vb{k}_\perp}^{2(G)},Ej} =0.
       \label{eq:expansion-coefficients-infinite-timeb}
    \end{align}
\end{subequations}

We can now define the total number of classical photons emitted by the charge (as seen by inertial observers) as
\begin{equation}
    N_{\text{cU}} 
    \equiv 
    \gkg{KRj,KRj}
    \equiv
    \| KRj \|_{\text{gKG}},
    \label{eq:classical-number-unruh-photons-1}
\end{equation}
where $KRj_a$ is the (inertial) positive-energy part of the retarded solution given in Eq.~\eqref{eq:field-eq}:
\begin{equation}
    KRj_a
    =
    -
    \sum_{\sigma,\kappa}
    \int_0^{\infty} \dd\omega \!
    \int_{\mathbb{R}^2} \dd^2\vb{k}_\perp
    \gkg{W^{\sigma(\kappa)}_{\omega\vb{k}_\perp}, Ej}
    W^{\sigma(\kappa)}_{\omega\vb{k}_\perp}{}_a .
        \label{eq:positive-energy-of-retarded-classical-solution}
\end{equation}
By using the orthonormality of the Unruh modes, we can write $ N_{\text{cU}} $ explicitly in terms of the coefficients of the expansion~\eqref{eq:orthogonality-EM-Unruh-modes} as   
\begin{equation}
    N_{\text{cU}}
    =
    \sum_{\sigma,\kappa}
    \int_0^\infty\dd\omega
    \int_{\mathbb{R}^2} \dd^2\vb{k}_\perp 
    \qty|\gkg{
            W^{\sigma(\kappa)}_{\omega\vb{k}_\perp} , Ej
    }|^2
    ,
    \label{eq:classical-number-unruh-photons-2}
\end{equation}
which enables us to interpret $ |\gkg{W^{\sigma(\kappa)}_{\omega\vb{k}_\perp} , Ej}|^2 $ as the number of photons associated to the Unruh mode $ \sigma $, per
polarization mode $\kappa $, per transverse momentum $ \vb{k}_\perp $ and Rindler energy $ \omega $. To avoid divergences, we will deal with the number of Unruh photons per transverse momentum
\begin{equation}
    \frac{\dd N_{\text{cU}}}{\dd^2\vb{k}_\perp} 
   \equiv 
    \sum_{\sigma,\kappa}
    \int_0^\infty\dd\omega
    \,
    \abs{
        \gkg{
            W^{\sigma(\kappa)}_{\omega\vb{k}_\perp} , Ej
        }
    }^2.
    \label{eq:classical-number-unruh-photons-3}
\end{equation}
In the limit $ T\to\infty $, we can use Eqs.~\eqref{eq:expansion-coefficients-infinite-timea} and~\eqref{eq:expansion-coefficients-infinite-timeb} in Eq.~\eqref{eq:classical-number-unruh-photons-3}, together with the property $W^{2(2)}_{\omega\vb{k}_\perp\ a} = W^{1(2)}_{-\omega\vb{k}_\perp\ a} $ to write
\begin{align}
    \frac{ \dd N_{\text{cU}}}{\dd^2\vb{k}_\perp} 
    &=
    \frac{q^2}{4 \pi^3 a} 
    \abs{K_1(k_\perp/a)}^2
    T_{\text{tot}},
    \label{eq:classical-number-unruh-photons-5}
\end{align}
where we have used that $ T_{\text{tot}} = 2\pi \delta(\omega)|_{\omega=0} $. 
This gives exactly the total rate rate of emission and absorption of zero-energy Rindler photons in the Unruh thermal bath seen by an accelerated observer, and the rate of emission detected by an inertial one, as computed in Ref.~\cite{PhysRevD.46.3450} using tree-level QFT.
It interesting to note how this (classical) radiation is built from zero-Rindler-energy Unruh modes, as can be seen explicitly from the form of the amplitudes in Eq.~(\ref{eq:expansion-coefficients-infinite-timea}).

Now, let us explicitly compute the expansion \eqref{eq:Unruh-mode-expansion} and show it reduces to the usual and well known solution for the electromagnetic 4-potential \eqref{eq:Born-potential-inertial} in the asymptotic future. For this purpose, let us focus on region EDKU in Fig.~\ref{fig:diagram}, the so called Expanding Degenerate Kasner Universe (the region where $t>|z|$). In the limit $T\to\infty $, we can see from Eqs.~\eqref{eq:expansion-coefficients-infinite-timea} and~~\eqref{eq:expansion-coefficients-infinite-timeb} that the only modes that couple with our current are $
W_{\omega\vb{k}_\perp\ a}^{\sigma(2)} $. As a result, by using again that $ W^{2(2)}_{\omega\vb{k}_\perp\ a} = W^{1(2)}_{-\omega\vb{k}_\perp\ a} $ to perform the integration in $\omega$,  Eq.~\eqref{eq:Unruh-mode-expansion} can be expressed as
\begin{align}
    Rj_a
    &=
    - \frac{q}{\sqrt{2a\pi^2}}
    \int_{\mathbb{R}^2} \dd^2\vb{k}_\perp
    \qty[
         i K_1(k_\perp/a) 
            W_{0\vb{k}_\perp}^{2(2)}{}_{a}
        +
        \text{c.c.}
    ]
    .\label{Rjexp0}
\end{align}

On the EDKU let us define the coordinates $ (\eta,\zeta,\vb{x}_\perp) $ by
\begin{align}
    t &= a^{-1} e^{a\eta} \cosh(a\zeta) ,
    &
    z &= a^{-1} e^{a\eta} \sinh(a\zeta) ,
    \label{eq:EDKU-rindler-coordinates}
\end{align}
while keeping the transverse position coordinates $ \vb{x}_\perp $ unaltered. In such coordinates, the vector Unruh mode
\eqref{eq:explicit-EM-unruh-modes-b} reads
\begin{equation*}
    W_{\omega\vb{k}_\perp}^{2(2)} {}_a
        =
        \frac{1}{k_\perp} 
        ( 
            \partial_\zeta w^{2}_{\omega\vb{k}_\perp}, 
            \partial_\eta w^{2}_{\omega\vb{k}_\perp},
            0,0 
        )
\end{equation*}
with the scalar Unruh mode, $w^{2}_{\omega\vb{k}_\perp}$,  having the rather simple form in the EDKU
\begin{equation} \label{w2EDKU}
    w^{2}_{\omega\vb{k}_\perp}
    =
    -
    \frac{i e^{\pi\omega/(2a)}}{\sqrt{32\pi^2a}}
    e^{i(\vb{k}_\perp\cdot\vb{x}_\perp + \omega\zeta)}
    H^{(2)}_{i\omega/a}(k_\perp e^{a\eta}/a)
    .
\end{equation}
From  Eq.~(\ref{w2EDKU}) we can see that $ \partial_\zeta w^{2}_{0\vb{k}_\perp} = 0 $ and, therefore, the only non-zero component of the 4-potential will be  
\begin{widetext}
    \noindent 
    \begin{align}
        Rj_\zeta (x)
        &=
        \frac{q e^{a\eta}}{8 \pi^2 a} 
        \int_{\mathbb{R}^2} \dd^2\vb{k}_\perp
        \qty(
            K_1(k_\perp/a)
            H^{(2)}_{1}(k_\perp e^{a\eta}/a)
            e^{i\vb{k}_\perp\cdot\vb{x}_\perp}
            +
            \text{c.c.}
        )
        ,
        \label{Rj1}
    \end{align}
    where we have used the identity $ {H^{(2) \prime}_{0}}(z) = -H^{(2)}_{1} (z) $ for the Hankel functions. To carry out the rest of the integration let us define polar coordinates for the transverse momentum vector by $\vb{x}_\perp \equiv   ( x_\perp \cos\varphi , x_\perp \sin\varphi) $ and  $
    \vb{k}_\perp \equiv \boldsymbol( k_\perp \cos(\varphi+\vartheta) , k_\perp\sin(\varphi+\vartheta) \boldsymbol),$ with $ {k}_\perp > 0$ and $ 0 \leq
    \vartheta < 2\pi $. As a result, $ \vb{k}_\perp \cdot \vb{x}_\perp = {k}_\perp x_\perp \cos\vartheta $ and $ \dd^2\vb{k}_\perp = \dd\vartheta \dd{k}_\perp
    k_\perp $, so we can rewrite the nonzero component of the retarded solution in Eq.~\eqref{Rj1} as 
    \begin{equation*}
        Rj_\zeta (x)
        =
        \frac{q e^{a\eta}}{8 \pi^2 a} 
        \int_{0}^{\infty} k_\perp \dd{k}_\perp
       \left[
            K_1(k_\perp/a)
            H^{(2)}_{1}(k_\perp e^{a\eta}/a)
            \int_{0}^{2\pi} \dd\vartheta \,
                \exp(i k_\perp x_\perp \cos\vartheta)
            +
            \text{c.c.}
        \right].
    \end{equation*}
\end{widetext}
Here, we recognize one of the integral form of the Bessel function of order zero \cite{gradshteyn2014table}
\begin{equation*}
    J_0 (z) = 
    \frac{1}{2\pi}
    \int_{0}^{2\pi} \dd\vartheta \,
            \exp({i z \cos\vartheta})
    ,
\end{equation*}
and use the definition of the second Hankel functions $ H^{(2)}_{\nu}(z) = J_{\nu}(z) - i Y_{\nu}(z) $, to show that
\begin{equation}
    Rj_\zeta 
    =
    \frac{q e^{a\eta}}{2 \pi a} 
    \!
    \int_{0}^{\infty} \!
        k_\perp
        K_1(k_\perp/a)
        J_{1}(k_\perp e^{a\eta}/a)
        J_0 (k_\perp x_\perp) 
        \dd{k}_\perp \!
        .
\end{equation}
We can now apply the identity \cite{higuchi2017entanglement,Bailey1936}, 
\begin{multline}
    \int_0^\infty 
    K_1(\alpha \vartheta) J_1(i \beta \vartheta) J_0(\gamma \vartheta) \vartheta
    \dd\vartheta
    \\
    =
    \frac{i}{2\alpha\beta}
    \left[ 
        \frac{
            \alpha^2 + \beta^2 + \gamma^2
        }{
            \sqrt{(\alpha^2 + \beta^2 + \gamma^2)^2 - 4\alpha^2\beta^2}
        }
        -1
    \right]
        ,
    \label{eq:integration-identity}
\end{multline}
which holds for the $ \mathrm{Re}\,\alpha > 0 $, $\mathrm{Re}\,\beta = 0 $, and $ \gamma > 0  $. If we identify $ \alpha = a^{-1} $, $
\beta = - i e^{a\eta} /a $ and $\gamma = x_{\perp} $, it is immediate to find
\begin{equation}
    Rj_\zeta (x)
    = 
    - \frac{qa}{4\pi}
    \left[
        \frac{
            a(a^{-2} - a^{-2} e^{2 a \eta}  + {x_\perp}^2)
        }{
            2 \rho_0(x)
        }
        -1
    \right]
    .
\end{equation}
This 4-potential is gauge-equivalent to the Born solution in Eq.~ \eqref{eq:Born-potential-inertial}. To prove this, we can define a scalar function 
$ \Lambda(x) \equiv -q a(\zeta+\eta)/4\pi $ and apply the gauge transformation $ Rj_a \to Rj_a + \nabla_a \Lambda $ to obtain exactly the solution originally found
by Born, but written in Rindler coordinates. Again, it interest to note how the usual retarded solution~(\ref{eq:Born-potential-inertial}) is built entirely from zero-Rindler-energy Unruh modes, as can be seen from Eq.~(\ref{Rjexp0}).
\section{Unruh mode decomposition of the quantum potential}\label{secdecomposition-quantum}

Let us now analyze the quantum aspects of the radiation emitted by the charge and its relation with the Unruh thermal bath. To this end, we will focus on the quantum 4-potential operator $ \hat A^{a} $, defined as a solution to the field equation 
\begin{equation}
     \nabla_b\nabla^b \hat{A}^a = j_L^a \hat{\mathbb{I}} .
     \label{eq:quantumA}
\end{equation}.

We can write a this operator in different ways, depending on the boundary/initial conditions chosen. One suitable choice is
\begin{equation}
    \hat A_a = \hat{A}^{\text{in}}_a + {Rj_L}_a \hat{\mathbb{I}},
    \label{Ain}
\end{equation}
where $\hat{A}^{\text{in}}_a $ is the solution to the homogeneous field equation, 
\begin{equation}
   \nabla_b\nabla^b \hat{A}^{\text{in}}_a = 0,
   \label{homquantumAin}
\end{equation}
 and we recall that ${Rj_L}^a$ is the retarded solution associated with the current $j_L^a.$
As a result we can expand  $\hat{A}^{\text{in}}_a $ as
\begin{equation}
    \hat{A}^{\text{in}}_a (t,\vb{x}) = 
    \sum_j \qty(
        u_{(j) \, a }(t,\vb{x}) \hat{a}_{\text{in}}(u_{(j)}^*)
        +
        \text{H.c.}
    ),
    \label{eq:in-expansion}
\end{equation}
where $ \{ u_{(j)} \}_{j\in \mathfrak{J}} $, with $\mathfrak{J}$ being a suitable set of quantum numbers, is a complete  set of (Minkowski) positive-frequency modes.  We can define $ \ket*{0^M_{\text{in}}} $ as the state such that $ \hat{a}_{\text{in}}(u_{(j)}^*) \ket*{0^M_{\text{in}}} = 0 $, for all $ j\in \mathfrak{J} $. As 
 $Rj_L^a$ vanishes in the asymptotic past (for instance, on the Cauchy surface $\Sigma_-$ in Fig.~\ref{fig:diagram}), one can interpret $\ket*{0^M_{\text{in}}}  $ as the vacuum state as seen by inertial observers in the asymptotic past. The Fock space describing particle states as seen by such observers is generated by the states
 \begin{equation}
    \ket*{ n_{j_1},n_{j_2},\cdots}_{\text{in}} 
    =
    \bigotimes_{j = j_1}^\infty 
        \frac{ [\hat{a}^{\text{in} \dagger}(u_{(j)})]^{n_j} }{ \sqrt{ n_{j}! } } 
        \ket*{0^M_{\text{in}}},
    \label{eq:in-state}
\end{equation}
where $n_j\in \mathbb{N}$ for each $j\in\mathfrak{J}$. 

Alternatively, we can also write a solution of Eq.~(\ref{eq:quantumA}) as 
\begin{equation}
    A_a = \hat{A}^{\text{out}}_a + A{j_L}_a \hat{\mathbb{I}} ,
    \label{Aout}
\end{equation}
 where $\hat{A}^{\text{out}}_a $ is a solution of the homogeneous field equation, 
 \begin{equation}
     \nabla_b\nabla^b
\hat{A}^{\text{out}}_a = 0, 
 \end{equation}
 and we recall that $Aj_L^a$ is the advanced solution associated with the current $j^a_L.$ Hence, we can expand $\hat{A}^{\text{out}}_a $ as 
\begin{equation}
    \hat{A}^{\text{out}}_a (t,\vb{x}) = 
    \sum_j \qty(
        v_{(j) \, a }(t,\vb{x}) \hat{a}^{\text{out}}(v_{(j)}^*)
        +
        \text{H.c.}
    ),
    \label{eq:out-expansion}
\end{equation}
where $ \{ v_{(j)} \}_{j\in \mathfrak{K}} $, with $\mathfrak{K}$ being a suitable set of quantum numbers, is any set of (Minkowski) positive-energy modes. We can then define $ \ket*{0^M_{\text{out}}} $ as the state
such that $ \hat{a}^{\text{out}}(v_{(j)}^*) \ket*{0^M_{\text{out}}} = 0 $, for all $ j\in \mathfrak{K}$. As $Aj_L^a$ vanishes in the asymptotic future (for example, in the Cauchy surface $\Sigma_+$ shown in Fig.~\ref{fig:diagram}), we can interpret it as the vacuum state seen by an inertial observer in the asymptotic future. We can also  construct the Fock space describing the particle states of the field by successive applications of the \emph{creation
operators} $\hat{a}^{\text{out} \dagger}(v_{(j)}) $ on $ \ket*{0^M_{\text{out}}}$, hence 
\begin{equation}
    \ket*{n_{j_1},n_{j_2},\cdots}_{\text{out}}
    =
    \bigotimes_{j = j_1}^\infty 
        \frac{ [\hat{a}^{\text{out} \dagger}(v_{(j)})]^{n_j} }{\sqrt{ n_{j}! }} 
        \ket*{0^M_{\text{out}}}.
    \label{eq:out-state}
\end{equation}

We can connect the in and out Fock spaces via the S-matrix~\cite{itzykson2012quantum}
\begin{equation}
    \hat S \equiv 
    \exp( 
        -i \int_{\mathbb{R}^4} \dd^4x \sqrt{-g}
        \hat{A}^{\text{out}}_a (t,\vb{x}) j_L^a(t,\vb x)
    ),
    \label{eq:s-matrix}
\end{equation}
which, in particular, relates the two vacua by
\begin{equation}
    \ket*{0^M_{\text{in}}} = \hat S \ket*{0^M_{\text{out}}}.
    \label{Svacuum}
\end{equation}
In order to compute this operator explicitly, let us expand the
out-field in terms of Unruh modes~(\ref{eq:explicit-EM-unruh-modes-d})
\begin{multline}
    \hat{A}^{\text{out}}_a =  
    \sum_{\sigma,\kappa}
    \int_0^\infty \dd \omega
    \int_{\mathbb{R}^2} \dd^2 \vb{k}_{\perp}
    \left[
        W_{\omega\vb{k}_\perp\ a}^{\sigma(\kappa)}
        \,
        \hat{a}_{\text{out}}(W_{\omega\vb{k}_\perp}^{\sigma(\kappa)*})
    \right.
    \\
    \left.
        +
        W_{\omega\vb{k}_\perp\ a}^{\sigma(\kappa) *}
        \,
        \hat{a}^\dagger_{\text{out}}(W_{\omega\vb{k}_\perp}^{\sigma(\kappa)})
    \right]
    .
    \label{eq:Unruh-expansion-out-field}
\end{multline}
In order to cast Eq.~(\ref{eq:Unruh-expansion-out-field}) in a more convenient form, let us first define the smearing of the quantum 4-potential with the classical current as
\begin{align}
    \hat{A}^{\text{out}}_a (j_L) 
    & \equiv 
     \int_{\mathbb{R}^4} \dd^4 x \sqrt{-g} \,
        \hat{A}^{\text{out}}_a (x) 
        j_L^a(x).
    \label{eq:out-field-smearing-w-source}
\end{align}
Next, by  using Eqs.~\eqref{eq:Unruh-expansion-out-field} and \eqref{eq:identity} we can write $ \hat{A}^{\text{out}}_a (j_L) $ as
\begin{multline}
    i  \hat{A}^{\text{out}}_a (j_L) 
    = 
    \sum_{\sigma,\kappa} 
    \int_0^\infty \dd \omega
    \int_{\mathbb{R}^2} \dd^2 \vb{k}_{\perp}
    \\
    \left[\gkg{W_{\omega\vb{k}_\perp}^{\sigma(\kappa)},Ej_L}^* 
        \,
        \hat{a}_{\text{out}}(W_{\omega\vb{k}_\perp}^{\sigma(\kappa)*})
    \right.
    \\
    -
    \left.
            \gkg{W_{\omega\vb{k}_\perp}^{\sigma(\kappa)},Ej_L}
        \,
        \hat{a}^\dagger_{\text{out}}(W_{\omega\vb{k}_\perp}^{\sigma(\kappa)})
    \right],
    \label{eq:out-field-smearing-w-source-exp-prod}
\end{multline}
This motivates the definitions of the creation and annihilation operators associated with the (inertial) positive energy part of the expansion as
\begin{multline}
    \hat{a}^\dagger_{\text{out}}(KEj_L) 
    \equiv \\
    \sum_{\sigma,\kappa} 
    \int_0^\infty \dd \omega
    \int_{\mathbb{R}^2} \dd^2 \vb{k}_{\perp}
    \gkg{W_{\omega\vb{k}_\perp}^{\sigma(\kappa)},Ej_L}
    \hat{a}^\dagger_{\text{out}}(W_{\omega\vb{k}_\perp}^{\sigma(\kappa)})
     \label{eq:tot-creation-operator}
\end{multline}
and 
\begin{multline}
    \hat{a}_{\text{out}}(KEj_L^*) 
    \equiv \\
    \sum_{\sigma,\kappa} 
    \int_0^\infty \dd \omega
    \int_{\mathbb{R}^2} \dd^2 \vb{k}_{\perp}
    \gkg{W_{\omega\vb{k}_\perp}^{\sigma(\kappa)},Ej_L}^* 
    \hat{a}_{\text{out}}(W_{\omega\vb{k}_\perp}^{\sigma(\kappa)^*})
    ,
    \label{eq:tot-annihilation-operator}
\end{multline}
respectively. By means of Eqs.~(\ref{eq:tot-creation-operator}) and~(\ref{eq:tot-annihilation-operator})  the S-matrix~\eqref{eq:s-matrix} can be cast as 
\begin{equation}
    \hat S = \exp(
        \hat{a}^\dagger_{\text{out}}(KEj_L) 
        - 
        \hat{a}_{\text{out}}(KEj_L^*)
    ).
    \label{eq:s-matrix-total-positive-parts}
\end{equation}
Now, by using the canonical commutation relation
\begin{align}
    \left[ \hat{a}_{\text{out}}(KEj_L^*), \hat{a}^\dagger_{\text{out}}(KEj_L) \right]
    &= \| KEj_L \|_{\text{gKG}}^2 \hat{\mathbb{I}},
    \label{eq:aux-comm2}
\end{align}
together with Zassenhaus formula 
\begin{equation}
    e^{\hat X + \hat Y} 
    = 
    e^{\hat X} e^{\hat Y} e^{-[\hat X, \hat Y]/2},
    \label{eq:Zassenhaus}
\end{equation}
where $\boldsymbol[\hat{X}, [\hat X, \hat Y]\boldsymbol]=\boldsymbol[\hat{Y}, [\hat X, \hat Y]\boldsymbol]=0$, we can rewrite Eq.~(\ref{eq:s-matrix-total-positive-parts}) as

\begin{equation}
    \hat S 
    =
    e^{- \| KEj_L \|_{\text{gKG}}^2/2}
    e^{
        \hat{a}^\dagger_{\text{out}}(KEj_L)
    }
    e^{
        -\hat{a}_{\text{out}}(KEj_L^*)
    }
    .
    \label{eq:potential-S}
\end{equation}

Note that the above S-Matrix is completely determined as we have already found the expansion amplitudes $\gkg{W_{\omega\vb{k}_\perp}^{\sigma(\kappa)},Ej_L}$, as well as the norm $\|KEj_L\|_{\text{gKG}}$ [which gives the classical number of
photons $N_{\rm cU}$]  in Sec.~\ref{sec:decomposition-classical}. After we recover our physical current (i.e., take $L\rightarrow \infty$) and  take the limit where our charge accelerates forever, $T\rightarrow\infty$, the non-vanishing amplitudes are given by Eq.~\eqref{eq:expansion-coefficients-infinite-timea} and the norm by
\begin{align}\label{norm}
    \| KEj \|_{\text{gKG}}^2 
    &=
    \frac{q^2}{2 a \pi^2} 
    T_{\text{tot}}
    \int_0^\infty \dd k_{\perp}\,  k_{\perp}
    \abs{K_1(k_\perp/a)}^2,
\end{align}
where we we recall that $T_{\rm tot}\equiv 2\pi \delta(\omega)|_{\omega=0}$ and we are using $j$ to denote the physical current describing the charge accelerating forever.

Now, by using Eqs.~(\ref{eq:expansion-coefficients-infinite-timea}) and~(\ref{eq:expansion-coefficients-infinite-timeb}) in Eqs.~(\ref{eq:tot-creation-operator}) and~(\ref{eq:tot-annihilation-operator}), together the fact that $W_{\omega {\bf k}_\perp}^{2 (\kappa)}=W_{-\omega {\bf k}_\perp}^{1 (\kappa)}$, we can cast $\hat{a}_{\text{out}}(KEj^*)$ and $\hat{a}^\dagger_{\text{out}}(KEj)$ as 

\begin{equation}\label{zeroa}
    \hat{a}_{\text{out}}(KEj^*)=\frac{iq}{\sqrt{2a}\pi}\int_{\mathbb{R}^2}\dd^2{\bf k}_\perp K_1\left(k_\perp/a\right) \hat{a}_{\rm out}\left(W^{2 (2) *}_{0 {\bf k}_\perp}\right)
\end{equation}
and 

\begin{equation} \label{zeroadagger}
    \hat{a}^\dagger_{\text{out}}(KEj)=\frac{-iq}{\sqrt{2a}\pi}\int_{\mathbb{R}^2}\dd^2{\bf k}_\perp K_1\left(k_\perp/a\right) \hat{a}^\dagger_{\rm out}\left(W^{2 (2)}_{0 {\bf k}_\perp}\right),
\end{equation}
respectively. If we now use Eqs.~(\ref{eq:potential-S})-(\ref{zeroadagger}) in Eq.~(\ref{Svacuum}),  we can write the in-vacuum $|0^M_{\rm in}\rangle$  as 
\begin{multline}
    \ket*{0^M_{\text{in}}} 
    =
    \bigotimes_{\vb{k}_\perp \in \mathbb{R}^2}
    \exp\left(
        - \frac{q^2 
        |K_1(k_\perp/a)|^2 T_{\text{tot}}}
        {8a\pi^3}\right)
    \\ \times
    \exp( 
        \frac{i q}{2\pi} \sqrt{\frac{2}{a}}
        K_1(k_\perp/a)\,
        \hat{a}^\dagger_{\text{out}}(W_{0\vb{k}_\perp}^{2(2)})
    )
    \ket*{0^M_{\text{out}}}
    ,
    \label{eq:in-vacuum-T-inf}
\end{multline}
which explicitly shows that the radiation emitted by the charge in the asymptotic future is built entirely by zero-Rindler-energy Unruh photons when the field is initially in the Minkowski vacuum. Moreover, Eq.~(\ref{eq:in-vacuum-T-inf}) is a (multi-mode) coherent state with respect to out-Unruh modes. To see this, let us show that $\ket*{0^M_{\text{in}}}$ is an eigenstate of $\hat{a}_{\text{out}}(W_{\omega \vb{k}_\perp}^{\sigma (\kappa) * })$ eigenvalue $\langle W_{\omega \vb{k}_\perp}^{\sigma (\kappa)}, Ej  \rangle$. To this end, let us first note that Eq.~\eqref{eq:potential-S} is equivalent to
\begin{equation}
    \ket*{0^M_{\text{in}}} 
    =
    e^{- \| KEj \|_{\text{gKG}}^2/2}
    e^{ \hat{a}^\dagger_{\text{out}}(KEj) }
    \ket*{0^M_{\text{out}}}
    ,
    \label{eq:in-vacuum-div}
\end{equation}
where we recall that $\|KEj\|_{\text{gKG}}$ and $\hat{a}^\dagger_{\text{out}}(KEj)$ are given in Eqs.~(\ref{norm}) and~(\ref{zeroadagger}), respectively. Now, let us apply $\hat{a}_{\text{out}}(W_{\omega\vb{k}_\perp}^{\sigma(\kappa)^*}) $ to Eq.~\eqref{eq:in-vacuum-div} to find 
\begin{widetext}
    \begin{equation}
        \hat{a}_{\text{out}}\left(W_{\omega\vb{k}_\perp}^{\sigma(\kappa)^*}\right)
        \ket*{0^M_{\text{in}}} 
        =
        e^{- \| KEj \|_{\text{gKG}}^2/2}
        e^{
            \hat{a}^\dagger_{\text{out}}(KEj)
        }
        \left(
            e^{
                -\hat{a}^\dagger_{\text{out}}(KEj)
            }
            \hat{a}_{\text{out}}(W_{\omega\vb{k}_\perp}^{\sigma(\kappa)^*})
            e^{
                \hat{a}^\dagger_{\text{out}}(KEj)
            }
        \right) 
        \ket*{0^M_{\text{out}}}
        .
        \label{eq:eigenvalue-1}
    \end{equation}
\end{widetext}
By using the identity 
\begin{equation}
    e^{-\hat X} \hat Y e^{\hat{X}} 
    =
    \hat Y 
    -
    [\hat X, \hat Y]
    +
    \frac{1}{2!} \boldsymbol[
        \hat X, [\hat X, \hat Y]
    \boldsymbol]
    +
    \ldots, 
    \label{eq:expBKH}
\end{equation}
and the canonical commutation relation 
\begin{equation}
    \left[ \hat{a}_{\text{out}}\left(W_{\omega\vb{k}_\perp}^{\sigma(\kappa)^*}\right),\hat{a}^\dagger_{\text{out}}\left(KEj\right) \right]
    =
     \gkg{W_{\omega \vb{k}_\perp}^{\sigma(\kappa)},Ej}
    \hat{\mathbb{I}},
    \label{eq:comm-aux2}
\end{equation}
one can see that Eq.~\eqref{eq:eigenvalue-1} can be rewritten as 
\begin{align}
    \hat{a}_{\text{out}}(W_{\omega\vb{k}_\perp}^{\sigma(\kappa)^*})
    \ket*{0^M_{\text{in}}}
    &= 
    \gkg{W_{\omega \vb{k}_\perp}^{\sigma(\kappa)},Ej} 
    \ket*{0^M_{\text{in}}}
    ,
    \label{eq:in-vacuum-is-eigenstate}
\end{align}
which proves the in-vacuum is an eigenstate of the out-annihilation operator associated to \emph{any} Unruh mode, i.e., a multimode coherent state.

The previous result gives us a straightforward way to show that the expectation value of the out-potential \eqref{eq:Unruh-expansion-out-field} in the in-vacuum in the asymptotic future is given by 
\begin{equation}
    \bra*{0^M_\text{in}}
    \hat{A}^{\text{out}}_a (x) 
    \ket*{0^M_\text{in}}
    = -Rj(x) 
    ,
    \label{eq:expectation-value-out-potential}
\end{equation}
where we have used that $Ej=-Rj$ in the asymptotic future. This implies that the out-field has an expectation value given by
\begin{align}
    \bra*{0^M_\text{in}}
    \hat{F}^{\text{out}}_{ab}  
    \ket*{0^M_\text{in}}
    &= 
    - RF_{ab},
    \label{eq:expectation-faraday}
\end{align}
where $ RF_{ab} = \nabla_a Rj_b - \nabla_b Rj_a $ is the classical retarded Faraday tensor. 

The coherent state structure of~\eqref{eq:in-vacuum-is-eigenstate} is also useful to find the number of photons radiated by the charge as seen by inertial observers in the asymptotic future. For this purpose, let us note that Eq.~\eqref{eq:in-vacuum-is-eigenstate} implies that 
\begin{equation}
    \bra*{0^M_\text{in}}
    \hat N_{\text{out}}(W^{\sigma(\kappa)}_{\omega \vb k_\perp})
    \ket*{0^M_\text{in}}
    =
    \qty|\gkg{W_{\omega \vb{k}_\perp}^{\sigma(\kappa)},Ej}|^2 ,
    \label{eq:quant-number-op-expectation-value}
\end{equation}
where we used the number operator
\begin{equation}
    \hat
N_{\text{out}}(W^{\sigma(\kappa)}_{\omega \vb k_\perp}) \equiv \hat a^\dagger_{\text{out}}(W^{\sigma(\kappa)}_{\omega \vb k_\perp}) \hat
a_{\text{out}}(W^{\sigma(\kappa)*}_{\omega \vb k_\perp}).
\end{equation} 
The result of Eq.~\eqref{eq:quant-number-op-expectation-value} coincides with our interpretation of the classical number of Unruh photons~\eqref{eq:classical-number-unruh-photons-2}. We can now  add for all polarizations $ \kappa $, Unruh modes $ \sigma $, and integrate them with respect to Rindler energy $ \omega $ to obtain the quantum number of
photons per transverse momentum $ \vb k_\perp $
\begin{align}
    \frac{\dd N^{\text{out}}_{\text{qU}} }{\dd^2 \vb{k}_\perp}
        &\equiv 
        \sum_{\sigma,\kappa} 
        \int_0^\infty \dd\omega 
            \bra*{0^M_\text{in}}
            \hat N_{\text{out}}(W^{\sigma(\kappa)}_{\omega \vb k_\perp})
            \ket*{0^M_\text{in}}
    \nonumber
        \\&= 
        \sum_{\sigma,\kappa} 
        \int_0^\infty \dd{\omega} 
        \qty|\gkg{W_{\omega \vb{k}_\perp}^{\sigma(\kappa)},Ej}|^2, 
\end{align}
which we can compute explicitly in the $ T\to\infty $ case by using Eq.~\eqref{eq:classical-number-unruh-photons-5}  yielding 
\begin{equation}
    \frac{\dd N^{\text{out}}_{\text{qU}} }{\dd^2 \vb{k}_\perp}
    =
        \frac{q^2}{4 a \pi^2} |K_1(k_\perp/a)|^2 T_{\text{tot}}
    .
    \label{eq:number-photons}
\end{equation}
This is consistent with our previous result in the classical case, and thus, also with the result obtained by~\citeauthor{PhysRevD.46.3450}~\cite{PhysRevD.46.3450} in the tree-level QFT calculations using the Unruh thermal bath.

To conclude, we can now study the \emph{normal ordered} stress-energy tensor for the out-field,
\begin{equation}
    : \!\! T^{ab}_{\text{out}} \!\!: \equiv 
    g_{c d} \!
    : \!\! F^{a c}_{\text{out}} F^{b d}_{\text{out}} \!\! :
    -
    \frac{1}{4} 
    g^{ab}\! 
    : \!\! F^{c d}_{\text{out}} F_{c d}^{\text{out}} \!\! :.
    \label{eq:normal-ordered-stress-energy}
\end{equation}
From Eq.~\eqref{eq:in-vacuum-is-eigenstate} it is simple to prove that $ : \!\! F^{a b}_{\text{out}} F^{c d}_{\text{out}} \!\! : $  will coincide with the product between two retarded classical Faraday tensors:
\begin{equation}
    \bra*{0^M_{\text{in}}}
    : \!\! F^{a b}_{\text{out}} F^{c d}_{\text{out}} \!\! :
    \ket*{0^M_{\text{in}}}
    = RF^{a b} RF^{c d},
\end{equation}
and, therefore, the expectation value of the stress-energy momentum tensor \eqref{eq:normal-ordered-stress-energy} is given by 
\begin{multline}
    \bra*{0^M_{\text{in}}}
    : \!\! T^{ab}_{\text{out}} \!\! : 
    \ket*{0^M_{\text{in}}}
    \\=
    g_{c d}
    RF^{a c} RF^{b d}
    -
    \frac{1}{4} 
    g^{ab}
    RF^{c d} RF_{c d},
    \label{eq:exp-value-normal-ordered-stress-energy}
\end{multline}
which is exactly the classical value of the stress-energy tensor for the out-field. This implies, in particular, that the energy flux integrated over a large sphere $S^2$ in the asymptotic future gives 
\begin{equation}
     \int_{S^2} \dd S^a \bra*{0^M_{\text{in}}}
    : \!\! T_{ab}^{\text{out}} \!\! : \left(\partial_t\right)^b
    \ket*{0^M_{\text{in}}}=\frac{2}{3} \left(\frac{q^2a}{4 \pi}\right),
\end{equation}
which is the usual Larmor formula. Here, $\dd S^a$ is the vector-valued volume element on the sphere and $(\partial_t)^a$ is the Killing field associated with the global inertial congruence.


\section{Final discussion} \label{sec:conclusions}

In this work we have studied the classical and quantum radiation emitted by a classical charge. In the classical realm, we have shown, by extending the definition of the well known Unruh modes used in scalar electrodynamics to the vector case, that only zero-Rindler-energy  Unruh modes contribute to the decomposition for the retarded potential and the corresponding amplitudes for our classical current in the limit where it accelerates for an infinite amount of proper time. We have also shown that the number of (classical) photons recover the usual results for the photon emission computed using tree-level quantum field theory~\cite{PhysRevD.46.3450}. 

For the quantum analysis, we have computed the S-Matrix connecting the quantum field theory construction made by inertial observers in the asymptotic past and future.  We have found that  if the state
is prepared in the (Minkowski) vacuum associated with the past inertial observers, the future ones will detect this as a multi-mode coherent state built entirely from zero-Rindler-energy Unruh modes in the limit of infinite proper acceleration. In addition, the expectation values of the 4-potential, Faraday field tensor, and stress-energy tensor are in agreement with the corresponding classical counterparts that arise from the retarded solution described in the classical analysis.

As a result, we were able to extend the analysis made previously by one of the authors of the (classical and quantum aspects of the) radiation emitted  by  a scalar source coupled to a real scalar field to the more realistic case of a classical charge  coupled to the electromagnetic field. This enabled us to settle the two puzzling aspects of the interplay between the Unruh effect and Larmor radiation: {\bf (1)} How the (quantum) Unruh effect can be codified in the classical Larmor radiation and {\bf (2)} the key role played by the zero-energy Rindler modes in such context. 

\appendix 
\section{Derivation of Eq.~(\ref{eq:identity})}
\label{appA}

Let us prove the validity of the identity written in Eq.~(\ref{eq:identity}). For this purpose, let $( \mathcal{M}, g_{ab} )$  be a 4-dimensional globally hyperbolic spacetime  such that $ R_{ab}=0 $. Let us consider solutions $\tilde{A}^a$ of the homogeneous electromagnetic equation~(\ref{eq:homogeneous-EM}) and let $j^a$ be a compactly supported 4-current in $( \mathcal{M}, g_{ab} )$. Hence, we can choose a Cauchy surface $ \Sigma_- $ in the causal past of the support of $ j^a $ which will enable us to write the functional
\begin{align}
    I[\tilde A,j] \equiv 
    \int_{\mathcal{M}} \dd^4 x \sqrt{-g} \, \tilde{A}_a^* j^a 
    \label{eq:integral1A}
\end{align}
as
\begin{equation}
    I[\tilde A,j]= \int_{J^+(\Sigma_-)} \dd^4 x \sqrt{-g} \, \tilde{A}_a^* j^a,
        \label{eq:integral1A1}
\end{equation}
where $ J^+(\Sigma_-) $ denotes the causal future of $ \Sigma_- $. As the advanced solution $Aj^a$ satisfies Eq.~\eqref{eq:adv-solution}, we can cast Eq.~(\ref{eq:integral1A}) as
\begin{equation}
    I[\tilde A,j] 
    =
    \int_{J^+(\Sigma_-)} \dd^4 x \sqrt{-g} \, 
        \tilde{A}_a^*
        (\nabla_b \nabla^b Aj^a).
        \label{eq:integralA2}
\end{equation}
Now, the integrand of Eq.~(\ref{eq:integralA2}) can be cast as 
\begin{equation}
    \tilde{A}_a^* 
    (\nabla_b \nabla^b Aj^a) 
    = 
    \nabla_b 
    ( \tilde{A}_a^* \tensor\nabla{}^b Aj^a )
    \label{eq:derivs1A}
\end{equation}
which, from the properties of the covariant derivative and the definition of the current $\Xi^b[\tilde{A},Aj] $ given in Eq.~\eqref{eq:genKG-EM-current}, can be written as
\begin{equation}
    \tilde{A}_a^* 
    (\nabla_b \nabla^b Aj^a) 
    = 
    i \nabla_a \Xi^a[\tilde{A},Aj]
    + 2 R_{ab}
        \tilde{A}^{a*} Aj^{b}.
        \label{derivA2}
\end{equation}
Here, we have used that one can write Eq.~(\ref{eq:genKG-EM-current}) as 
\begin{equation}
    \Xi^b[\tilde{A},Aj] 
    = 
    2i \nabla_a (\tilde{A}^{[a*} Aj^{b]}) 
    - 
    i \tilde{A}_a^* {\stackrel{\leftrightarrow}{\nabla}}{}^b Aj^a
    ,
    \label{eq:current-1A}
\end{equation}
as well as $2\nabla_{[a} \nabla_{b]}A^c\equiv -R_{abd}{}^c A^d$ and $R_{ab}\equiv  R_{acb}{}^c$~\cite{Wald84}.  

If we now use Eqs.~(\ref{eq:derivs1A}) and~(\ref{derivA2}) in Eq.~(\ref{eq:integralA2}) together with the fact that we are dealing with Ricci-flat spacetimes ($R_{ab}=0$), we can write $I[\tilde{A}, j]$ using Gauss theorem as 
\begin{equation}
I[\tilde{A}, j]=i\int_{\Sigma_-} \dd \Sigma_a \Xi^a[\tilde{A},Aj]. 
    \label{intermgKG}
\end{equation}
Now, by using  Eqs.~(\ref{eq:gen-KG-prod}) and~\eqref{eq:integral1A} and noting that $Aj=Ej$ on $\Sigma_-$, since $Rj^a=0$ in this region, Eq.~(\ref{intermgKG}) can be cast as
 \begin{equation}
    \gkg{\tilde{A},Ej} 
    = 
    -i \int_{\mathcal{M}} \dd^4 x \sqrt{-g} \, \tilde{A}_a^* j^a,
    \label{eq:identityA}
\end{equation}
which is 
exactly the form of Eq.~\eqref{eq:identity}, as we wanted to prove.

\section{Compactified Current Calculations}
\begin{widetext}
\label{appB2}
We need to find the amplitudes $\mathcal{A}^{\sigma(\kappa)}_{\omega\mathbf{k}_\perp}$ and $\mathcal{B}^{\sigma(\kappa)}_{\omega\mathbf{k}_\perp}$ in order to determine the form of the corrections $ \mathcal{I}^{\sigma(\kappa)}_{\omega \vb{k}_\perp}$ appearing in the expansion coefficients defined in
Eq.~(\ref{eq:expansion-coefficients-finite-time}). For simplicity we shall only compute these for $ \kappa =2 $, as the amplitudes corresponding to the pure
gauge mode labeled with $ \kappa=G $ can be omitted. To this end, let us first note that we can use the fact that $ \Delta_+(z,-t) = \Delta_-(z,t) $ to rewrite
Eqs.~\eqref{eq:A.correction} and~\eqref{eq:B.correction} as 
\begin{align}
    \mathcal{A}^{\sigma(2)}_{\omega\mathbf{k}_\perp}
    & = 
    \int_{a^{-1} \sinh(a T)}^{L} \mathrm{d} t \int_{\mathbb{R}}\mathrm{d} z
    \left[
        W_{\omega\mathbf{k}_\perp}^{\sigma(2) * } {}_{t}(t,0,0,z) 
    +
    W_{\omega\mathbf{k}_\perp}^{\sigma(2) *}{}_{t}(-t,0,0,z) \right]
    \Delta_-(z,t)
    ,
    \label{eq:A.correction-aux1}
    \\
    \mathcal{B}^{\sigma(2)}_{\omega\mathbf{k}_\perp}
    & =  \int_{a^{-1} \sinh(a T)}^{L} \mathrm{d} t \int_{\mathbb{R}}\mathrm{d} z 
    \left[
        W_{\omega\mathbf{k}_\perp}^{\sigma(2) * } {}_{z}
            (t,0,0,z) 
    -  
    W_{\omega\mathbf{k}_\perp}^{\sigma(2) *}{}_{z} (-t,0,0,z) 
    \right]
    \Delta_-(z,t).
    \label{eq:B.correction-aux1}
\end{align}
Next, we can use the integral form of the scalar Unruh modes~\cite{higuchi2017entanglement}
\begin{equation*}
    w^{\sigma}_{\omega \mathbf{k}_\perp} =
    \frac{e^{i \mathbf{k}_\perp \cdot \mathbf{x}_\perp }}{(2\pi)^2 \sqrt{2a} }
    \int_{-\infty}^{\infty} \dd \vartheta \, 
    e^{i (-1)^\sigma \vartheta\omega/a} 
    \exp[
        i k_\perp (z \sinh\vartheta - t \cosh\vartheta)
    ],
\end{equation*}
to find integral expressions for the components of the vector Unruh modes we need [see Eq.~\eqref{eq:explicit-EM-unruh-modes-b}]:
    \begin{gather}
        W^{\sigma(2) *}_{\omega \mathbf{k}_\perp} {}_t = 
        -\frac{i e^{-i \mathbf{k}_\perp \cdot \mathbf{x}_\perp }}{(2\pi)^2 \sqrt{2a} }
        \int_{-\infty}^{\infty} \dd \vartheta \, 
        e^{-i (-1)^\sigma \vartheta\omega/a} \sinh\vartheta
        \exp[
            -i k_\perp (z \sinh\vartheta - t \cosh\vartheta)
        ],\label{intvecWa}
        \\
        W^{\sigma(2) *}_{\omega \mathbf{k}_\perp} {}_z = 
        +\frac{i e^{-i \mathbf{k}_\perp \cdot \mathbf{x}_\perp }}{(2\pi)^2 \sqrt{2a} }
        \int_{-\infty}^{\infty} \dd \vartheta \, 
        e^{-i (-1)^\sigma \vartheta\omega/a} \cosh\vartheta
        \exp[
            -i k_\perp (z \sinh\vartheta - t \cosh\vartheta)
        ].\label{intvecWb}
    \end{gather}
    Now, by applying Eqs.~(\ref{intvecWa}) and~(\ref{intvecWb}) into Eqs.~\eqref{eq:A.correction-aux1} and~\eqref{eq:B.correction-aux1} we find that
    \begin{gather}
        \mathcal{A}^{\sigma(2)}_{\omega\mathbf{k}_\perp}
         = -\frac{i}{2\pi^2 \sqrt{2a} }
        \int_{a^{-1} \sinh(a T)}^{L} \mathrm{d}t 
        \int_{-\infty}^{\infty} \mathrm{d}z
            \int_{-\infty}^{\infty} \dd \vartheta \, 
                e^{-i (-1)^\sigma \vartheta\omega/a} \sinh\vartheta \,
                e^{ -i k_\perp z \sinh\vartheta} 
                \cos (k_\perp t \cosh\vartheta)
        \Delta_-(z,t)
        ,
        \label{eq:A.correction-aux2}
        \\
        \mathcal{B}^{\sigma(2)}_{\omega\mathbf{k}_\perp}
         = -\frac{1}{2\pi^2 \sqrt{2a} }
        \int_{a^{-1} \sinh(a T)}^{L} \mathrm{d}t 
        \int_{-\infty}^{\infty} \mathrm{d}z
            \int_{-\infty}^{\infty} \dd \vartheta \, 
                e^{-i (-1)^\sigma \vartheta\omega/a} \cosh\vartheta \,
                e^{ -i k_\perp z \sinh\vartheta} 
                \sin (k_\perp t \cosh\vartheta)
        \Delta_-(z,t)
        .
        \label{eq:B.correction-aux2}
    \end{gather}
The integrals in the $z$ variable in the above expressions can be computed immediately yielding 
    \begin{gather}
        \mathcal{A}^{\sigma(2)}_{\omega\mathbf{k}_\perp}
         = -\frac{i}{2\pi^2 \sqrt{2a} }
        \int_{a^{-1} \sinh(a T)}^{L} \mathrm{d}t 
            \int_{-\infty}^{\infty} \dd \vartheta \, 
                e^{-i [(-1)^\sigma \vartheta\omega
                    +k_\perp \sech(aT) \sinh\vartheta]/a}
                \sinh\vartheta \,
                e^{ -i k_\perp t\tanh(aT)  \sinh\vartheta} 
                \cos (k_\perp t \cosh\vartheta)
        ,
        \label{eq:A.correction-aux3}
        \\
        \mathcal{B}^{\sigma(2)}_{\omega\mathbf{k}_\perp}
         = -\frac{1}{2\pi^2 \sqrt{2a} }
        \int_{a^{-1} \sinh(a T)}^{L} \mathrm{d}t 
            \int_{-\infty}^{\infty} \dd \vartheta \, 
                e^{-i [(-1)^\sigma \vartheta\omega
                    +k_\perp \sech(aT) \sinh\vartheta]/a}
                \cosh\vartheta \,
                e^{ -i k_\perp t\tanh(aT)  \sinh\vartheta} 
                \sin (k_\perp t \cosh\vartheta)
        .
        \label{eq:B.correction-aux3}
    \end{gather}
   For the sake of notation, let us define 
    \begin{gather}
        f^A_a(\vartheta,L,T) \equiv 
        \int_{a^{-1} \sinh(a T)}^{L} \mathrm{d}t \,
        e^{ -i k_\perp t\tanh(aT)  \sinh\vartheta} 
        \cos (k_\perp t \cosh\vartheta),
        \label{eq:A.correction-aux4}
        \\
        f^B_a(\vartheta,L,T) \equiv 
        \int_{a^{-1} \sinh(a T)}^{L} \mathrm{d}t \,
        e^{ -i k_\perp t\tanh(aT)  \sinh\vartheta} 
        \sin (k_\perp t \cosh\vartheta),
        \label{eq:B.correction-aux4}
    \end{gather}
    which we can use to cast $\mathcal{A}^{\sigma(2)}_{\omega\mathbf{k}_\perp}$ and $\mathcal{B}^{\sigma(2)}_{\omega\mathbf{k}_\perp}$ as 
    \begin{gather}
        \mathcal{A}^{\sigma(2)}_{\omega\mathbf{k}_\perp}
            = -\frac{i}{ 2\pi^2 \sqrt{2a} } 
            \int_{-\infty}^{\infty} \dd \vartheta \,
            e^{-i [(-1)^\sigma \vartheta\omega
                    +k_\perp \sech(aT) \sinh\vartheta]/a}
                f^A_a(\vartheta,L,T)
                \sinh\vartheta , 
        \label{eq:A.correction-aux5}
        \\
        \mathcal{B}^{\sigma(2)}_{\omega\mathbf{k}_\perp}
            = +\frac{i}{2\pi^2 \sqrt{2a} } 
            \int_{-\infty}^{\infty} \dd \vartheta \,
            e^{-i [(-1)^\sigma \vartheta\omega
                    +k_\perp \sech(aT) \sinh\vartheta]/a}
                f^B_a(\vartheta,L,T)
                \cosh\vartheta . 
        \label{eq:B.correction-aux5}
    \end{gather}
    We can see now from Eqs.~(\ref{eq:A.correction-aux5}) and~(\ref{eq:B.correction-aux5}) that, if we want to study the behavior of the physical current (recovered when $ L\to\infty $) and of the infinite acceleration proper-time  (i.e., $ T\to\infty $), it is enough to analyze $f_a(\vartheta,L,T) $ in such limits. 

    By integrating Eq.~(\ref{eq:A.correction-aux4}) we find that
    \begin{multline}
        f^A_a(\vartheta,L,T)
        =
        \frac{i \sinh\vartheta \tanh(aT) }{k_\perp[
            \sinh^2\vartheta \tanh^2(aT) - \cosh^2\vartheta
        ]}
        \left\{
            e^{
                i k_\perp L \tanh(aT) \sinh\vartheta
            }
            \left[
                \cos(k_\perp L \cosh\vartheta)
                +
                \frac{i \coth\vartheta}{\tanh(aT)}
                    \sin(k_\perp L \cosh\vartheta)
            \right]
        \right. 
        \\
        +\left.
            e^{
                i (k_\perp/a) \sinh(aT)\tanh(aT) \sinh\vartheta
            }
            \left[
                \cos\!\left( \frac{k_\perp}{a} \sinh(aT) \cosh\vartheta \right)
                +
                \frac{i \coth\vartheta}{\tanh(aT)}
                    \sin\!\left(\frac{k_\perp}{a} \sinh(aT) \cosh\vartheta \right)
            \right]
        \right\} ,
        \label{eq:A.correction-aux6}
    \end{multline}
    whilst Eq.~\eqref{eq:B.correction-aux4} yields 
    \begin{multline}
        f^B_a(\vartheta,L,T)
        =
        \frac{i \sinh\vartheta \tanh(aT) }{k_\perp[
            \sinh^2\vartheta \tanh^2(aT) - \cosh^2\vartheta
        ]}
        \left\{
            e^{
                i k_\perp L \tanh(aT) \sinh\vartheta
            }
            \left[
                \sin(k_\perp L \cosh\vartheta)
                -
                \frac{i \coth\vartheta}{\tanh(aT)}
                    \cos(k_\perp L \cosh\vartheta)
            \right]
        \right. 
        \\
        +\left.
            e^{
                i (k_\perp/a) \sinh(aT)\tanh(aT) \sinh\vartheta
            }
            \left[
                \sin\!\left( \frac{k_\perp}{a} \sinh(aT) \cosh\vartheta \right)
                -
                \frac{i \coth\vartheta}{\tanh(aT)}
                    \cos\!\left(\frac{k_\perp}{a} \sinh(aT) \cosh\vartheta \right)
            \right]
        \right\} .
        \label{eq:B.correction-aux6}
    \end{multline}

    Let us now take the limit $ L\to\infty$. To this end, we can use Eq.~\eqref{eq:limit} and the identity
    $
    \cos(\Omega L) = \sin\left[ \Omega \left(L + {\pi}/{2\Omega}\right)\right]
    $
    to see that, if $ \Omega \neq 0 $, we can set
    \begin{equation*}
        \lim_{L\to\infty} \cos(\Omega L) = \lim_{L\to\infty} \sin(\Omega L) 
        = 
        \pi\Omega \delta(\Omega) \overset{!}{=} 
        0.
    \end{equation*}
    As $ \cosh\vartheta \geq 1 $ for all values of $ \vartheta $, we have 
    \begin{multline}
        \lim_{L\to\infty} f^A_a(\vartheta,L,T)
        =
        \frac{i \sinh\vartheta \tanh(aT) }{k_\perp[
            \sinh^2\vartheta \tanh^2(aT) - \cosh^2\vartheta
        ]}
        e^{
                i (k_\perp/a) \sinh(aT)\tanh(aT) \sinh\vartheta
            }
        \\ \times \left[
            \cos\!\left( \frac{k_\perp}{a} \sinh(aT) \cosh\vartheta \right)
            +
            \frac{i \coth\vartheta}{\tanh(aT)}
                \sin\!\left(\frac{k_\perp}{a} \sinh(aT) \cosh\vartheta \right)
        \right],
        \label{eq:A.correction-aux7}
    \end{multline}
    \begin{multline}
        \lim_{L\to\infty} f^B_a(\vartheta,L,T)
        =
        \frac{i \sinh\vartheta \tanh(aT) }{k_\perp[
            \sinh^2\vartheta \tanh^2(aT) - \cosh^2\vartheta
        ]}
        e^{
                i (k_\perp/a) \sinh(aT)\tanh(aT) \sinh\vartheta
            }
        \\ \times \left[
            \sin\!\left( \frac{k_\perp}{a} \sinh(aT) \cosh\vartheta \right)
            -
            \frac{i \coth\vartheta}{\tanh(aT)}
                \cos\!\left(\frac{k_\perp}{a} \sinh(aT) \cosh\vartheta \right)
        \right].
        \label{eq:B.correction-aux7}
    \end{multline}
\end{widetext}
The same arguments can be used in the limit $ T\to\infty $ to see that 
\begin{equation}
    \lim_{T\to\infty} \lim_{L\to\infty} f^A_a(\vartheta,L,T)
    =
    \lim_{T\to\infty} \lim_{L\to\infty} f^B_a(\vartheta,L,T)
    =
     0,
    \label{eq:A.B.correction-aux8}
\end{equation}
which, by using Eq.~(\ref{eq:correction-smooth-mode2}),~(\ref{eq:A.correction-aux5}),  and~(\ref{eq:B.correction-aux5}) implies that  
\begin{equation}
    \lim_{T\to\infty} \lim_{L\to\infty} 
    \mathcal{I}^{\sigma(2)}_{\omega \vb{k}_\perp}
    = 0.
    \label{eq:correction-final-limits}
\end{equation}
The analysis for $ \mathcal{I}^{\sigma(G)}_{\omega \vb{k}_\perp} $ follows the same reasoning, as direct calculations will show that these depend on the same
integrals $ f^A_a(\vartheta,L,T) $ and $ f^B_a(\vartheta,L,T) $ defined in \eqref{eq:A.correction-aux4} and \eqref{eq:B.correction-aux4}.

\acknowledgments

We are thankful to Stephen Fulling and George Matsas for providing feedback on our manuscript.
This research was funded by grant \#2019/09401-4, S\~{a}o Paulo Research Foundation (FAPESP).

\bibliography{EMrad-UM}
\end{document}